\documentclass[useAMS,usenatbib]{mn2e}
\usepackage{psfig}
\usepackage{amssymb}
\usepackage{color}
\usepackage{enumerate}
\usepackage{rotating}
\usepackage{amsmath}
\usepackage{mathrsfs}
\usepackage{ragged2e}
\usepackage{hyperref}
\usepackage{amsmath}
\usepackage[export]{adjustbox}

\title[The physical origins of low-mass spin bias]
{The physical origins of low-mass spin bias}

\author[Tucci et al.]{
\parbox[t]{\textwidth}{
Beatriz Tucci$^{1}$\thanks{E-mail: beatriz.tucci@gmail.com}, Antonio D. Montero-Dorta$^{1,2}$, L. Raul Abramo$^{1}$, Gabriela Sato-Polito$^{3}$, M. Celeste Artale$^{4}$}
\vspace*{6pt} \\
$^1$ Departamento de F\'isica Matem\'atica, Instituto de F\'isica, Universidade de S\~ao Paulo, Rua do Mat\~ao 1371, CEP 05508-090, \\
S\~ao Paulo, Brazil\\
$^2$ Departamento de F\'isica, Universidad T\'ecnica Federico Santa Mar\'ia, Casilla 110-V, Avda. Espa\~na 1680, Valpara\'iso, Chile\\
$^3$ Department of Physics \& Astronomy, Johns Hopkins University, 3400 N. Charles St., Baltimore, MD 21218, USA \\
$^4$ Institut f\"ur Astro- und Teilchenphysik, Universit\"at Innsbruck, Technikerstrasse 25/8, 6020 Innsbruck, Austria 
\vspace{-0.4cm}
}

\date{Accepted ---. Received ---;in original form --- \vspace{-0.3cm}}

\def\simlt{\lower.5ex\hbox{$\; \buildrel < \over \sim \;$}}
\def\simgt{\lower.5ex\hbox{$\; \buildrel > \over \sim \;$}}
\usepackage{graphicx}
\usepackage{rotating}

\definecolor{red}{rgb}{1,0,0}

\begin{document}

\bibliographystyle{mnras}

\maketitle

\begin{abstract}

At $z=0$, higher-spin haloes with masses above $\log(\text{M}_{\text{c}}/h^{-1}\text{M}_\odot)\simeq 11.5$ have a higher bias than lower-spin haloes of the same mass. However, this trend is known to invert below this characteristic crossover mass, $\text{M}_{\text{c}}$. In this paper, we measure the redshift evolution and scale dependence of {\it{halo spin bias}} at the low-mass end and demonstrate that the inversion of the signal is entirely produced by the effect of  {\it{splashback haloes}}. These low-mass haloes tend to live in the vicinity of significantly more massive haloes, thus sharing their large-scale bias properties. We further show that the location of the redshift-dependent crossover mass scale $\text{M}_{\text{c}}(z)$ is completely determined by the relative abundance of splashbacks in the low- and high-spin subpopulations. Once splashback haloes are removed from the sample, the {\it{intrinsic}} mass dependence of spin bias is recovered. Since splashbacks have been shown to account for some of the assembly bias signal at the low-mass end, our results unveil a specific link between two different secondary bias trends: spin bias and assembly bias. 

\end{abstract}


\section{Introduction} 
\label{sec:intro}

At fixed halo mass, the large-scale clustering of dark-matter haloes has been shown to depend on a variety of internal halo properties, including age, concentration, spin, substructure content, or shape (e.g., \citealt{ShethTormen2004, gao2005, wechsler2006, gao2007, Wetzel2007, jing2007,li2008, Angulo2008, Faltenbacher2010, han2018,Chue2018, salcedo2018, Johnson2019, Sato-Polito2019, Montero-Dorta2020}). Investigating the physical origins of this {\it{secondary halo bias}} is relevant for our understanding of structure formation, our ability to construct realistic halo--galaxy connection models, and even for the extraction of cosmological information from galaxy maps (e.g., \citealt{Wu2008,Zentner2014,hearin2014, Hearin2016,Wechsler2018}).  

Among the secondary dependencies of halo clustering, {\it{halo spin bias}} is interesting because it connects the spatial distribution of haloes with their angular momentum, an important quantity in galaxy formation models. Until recently, the accepted view was that higher-spin haloes had higher bias than their lower-spin,  same-mass counterparts across the entire halo mass range \citep{gao2007,Bett2007,Faltenbacher2010, Lacerna2012, Villarreal2017, Lazeyras2017, Desjacques2018, Mao2018, salcedo2018}. However, \citet{Sato-Polito2019} used the high-resolution MultiDark numerical simulations\footnote{https://www.cosmosim.org/} to measure spin bias for very low halo masses, revealing a different picture. At $z=0$, low-spin haloes of $\log(\text{M}_{\text{vir}}/h^{-1}\text{M}_\odot)<11.5$ have actually higher bias than higher spin haloes, as subsequently confirmed by \citet{Johnson2019} using two independent simulations (Vishnu and Consuelo). The main goal of this paper is to shed light on the physical origins of this {\it{spin bias inversion}} and its evolution with redshift.  

So far, little work has been published on the physical causes of spin bias, since most of the effort has been placed on explaining the {\it{halo assembly bias}} trend (i.e., the dependence of halo clustering on age or concentration). Since different secondary bias trends might arise from similar physical mechanisms, investigating the causes of assembly bias can give us clues on the origins of spin bias. For the former, \cite{dalal2008} proposed a theory where this secondary dependence is caused by two intrinsically different effects that take place on different mass scales. At high-masses (M$\gg$M$_{*}$, where M$_{*}$ is the characteristic mass for which $\nu=\delta_\text{c}/\sigma(\text{M}_*,z)=1$), assembly bias can be explained from the statistics of Gaussian random fields. For low-mass haloes, on the other hand, the authors attribute the effect to a non-accreting subpopulation that lives in the vicinity of larger haloes, and thus share their large-scale bias. 

This environmental view of assembly bias has prompted several analyses, in which the arrested development of old low-mass haloes is connected to tidal interactions, hot environments, and to the so-called {\it{splashback haloes}} (e.g.,  \citealt{Wang2007, dalal2008, Hahn2009, Borzyszkowski2017, salcedo2018, Mansfield2020}). Also known as {\it{ejected subhaloes}}, splashbacks are distinct haloes at a given redshift that previously passed through the virial radius of a larger halo, experiencing extreme tides and showing early formation times (e.g., \citealt{Wang2009, Behroozi2014}). Other environment-related theories connect halo assembly bias with the cosmic web, claiming that the effect is a consequence of the anisotropy of the tidal environment \citep{Paranjape2018, Ramakrishnan2019, Ramakrishnan2020}.

For spin bias, most of the attempts to provide a plausible explanation have focused on the high-mass end (i.e., above the spin crossover). In this context, \cite{salcedo2018} showed little correlation between spin bias and the proximity to a significantly more massive halo, which seems to discard the massive-neighbour theory. \cite{Lacerna2012}, on the other hand, attributed the high bias of high-spin haloes to their location in the cosmic web, as material from filaments accreted by massive haloes (predominantly in high-density environments) can increase the haloes' angular momenta. This hypothesis seems to be in contradiction with the results of \cite{Johnson2019}, who in turn introduce the notion of ``twin'' bias, a tweak to the massive-companion argument by which high-spin haloes are slightly more likely to be found near other haloes of comparable mass. \cite{Johnson2019}, who actually address the entire mass range, further suggests that spin bias could in fact be described by a combination of twin bias and the contribution of other residual secondary dependencies. All the above works provide interesting information about spin bias, but a definite explanation for the inversion of the signal at the low-mass end is yet to be established.  

Motivated by previous results on assembly bias, we use several MultiDark boxes to analyse the effect that splashback haloes have on low-mass spin bias and its redshift evolution. We demonstrate that the inversion of the signal can be completely removed if splashbacks are excluded from the sample. This allows us to unveil the {\it{intrinsic}} scale and mass dependence of spin bias. 

The paper is organised as follows. The MultiDark boxes employed in this analysis are described in Section~\ref{sec:sims}. The redshift evolution of low-mass spin bias is presented in Section~\ref{sec:inversion}, while the effect of removing splashback haloes
is shown in Section~\ref{sec:splashback}. The main physical mechanism affecting splashback haloes is discussed in Section~\ref{sec:tidal}. Finally, in Section~\ref{sec:conc}
we present our main conclusions and discuss the implications of our results. Throughout this work, we assume the standard $\Lambda$CDM cosmology \citep{planck2014}, with parameters $h = 0.677$, $\Omega_m = 0.307$, $\Omega_{\Lambda} = 0.693$, $n_s = 0.96$, and $\sigma_8 = 0.823$.

\section{Simulations} 
\label{sec:sims}

We use the publicly available MultiDark suite of cosmological N-body simulations \citep{multidark2016}, from which we analyse three different simulation boxes: Very Small MultiDark Planck (VSMDPL), Small MultiDark Planck (SMDPL), and MultiDark Planck 2 (MDPL2). These boxes contain 3840$^3$ particles and span side lengths of 160, 400,  and 1000 $h^{-1}$Mpc, respectively. The halo catalogues that we employ were produced using the ROCKSTAR (Robust Overdensity Calculation using K-Space Topologically Adaptive Refinement) halo finder \citep{rockstar2013}, which identifies dark-matter haloes and their substructures, as well as tidal features. 

The halo virial mass, M$_{\text{vir}}$, is computed in ROCKSTAR assuming the virial threshold of \cite{Bryan1998}. We use the spin parameter, $\lambda$, as defined in \cite{bullock2001}, namely:
\begin{equation}
    \lambda = \frac{|J|}{\sqrt{2} \, \text{M}_{\text{vir}} \text{V}_{\text{vir}} \text{R}_{\text{vir}}},
\end{equation}
where $J$ is the halo angular momentum inside a sphere of radius $\text{R}_{\text{vir}}$ and mass $\text{M}_{\text{vir}}$, while $\text{V}_{\text{vir}}$ is its circular velocity at virial radius $\text{R}_{\text{vir}}$. Halo spin is therefore a dimensionless way of characterising the angular momentum of a dark matter halo. For a spherically symmetric object, the spin is basically the ratio between its angular velocity and the velocity needed for it to be rotationally supported. Importantly, we have checked that the distributions of the spin parameter are consistent across the MultiDark boxes used in this work.

The \citeauthor{bullock2001} $\lambda$ parameter has the advantage that it eliminates the energy dependence of the spin definition originally introduced by \cite{Peebles1969}, while reducing to the classical form when measured at the virial radius of a truncated singular isothermal halo. We refer the reader to \cite{Rodriguez-Puebla2016} for a discussion on the mass and redshift dependence of both spin parameters in MultiDark. 
\section{The redshift evolution of low-mass spin bias} 
\label{sec:inversion}

\begin{figure*}
    \centering
    \includegraphics[scale=0.58]{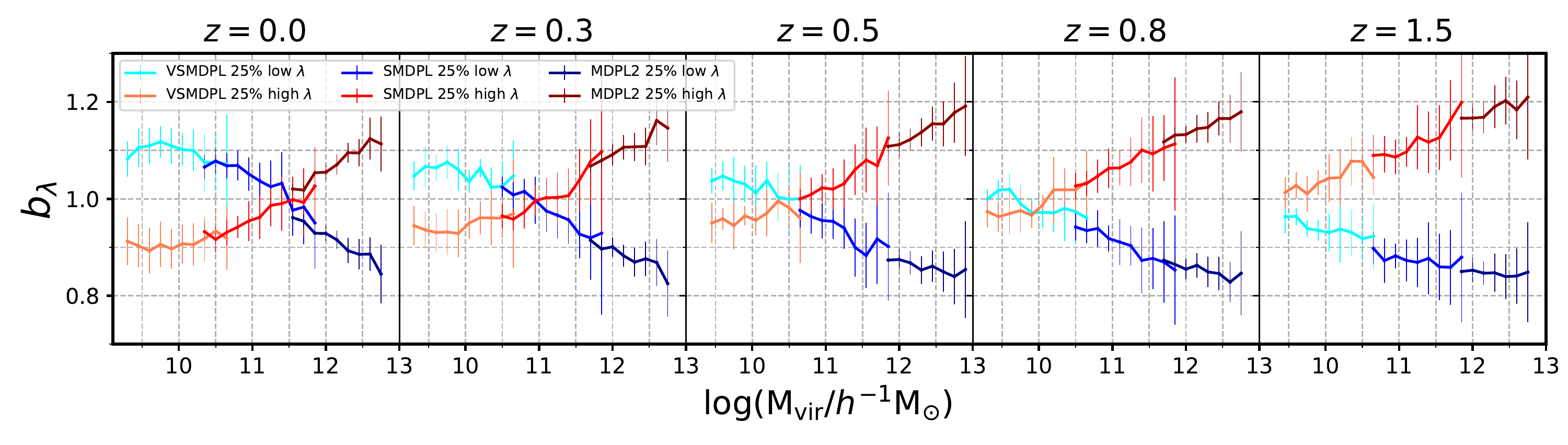}
    \caption{Redshift evolution of spin bias for the 25\% highest (red tones) and 25\% lowest (blue tones) spin subpopulations. Different tones represent different MultiDark boxes (from left to right: VSMDPL, SMDPL, and MDPL2, respectively). Error bars show the box-to-box variation computed from a set of subboxes (see text).}
    \label{spinbias}
\end{figure*}

The first measurement of spin bias for halo masses below $\log(\text{M}_{\text{vir}}/h^{-1} \text{M}_\odot )\simeq 12$ was presented in \citet{Sato-Polito2019}, using the $z=0$ SMDPL box. This measurement revealed for the first time the inversion of the signal below $\log(\text{M}_{\text{vir}}/h^{-1} \text{M}_\odot )\simeq 11.5$, where lower-spin haloes were shown to have higher bias than their higher-spin counterparts at fixed halo mass. This result was subsequently confirmed by \citet{Johnson2019} with MultiDark and two other independent simulations, Vishnu and Consuelo. In this section, we extend the analysis presented in \citet{Sato-Polito2019} by analysing the smaller VSMDPL box, which allows us to reach masses below $\log(\text{M}_{\text{vir}}/h^{-1} \text{M}_\odot )\simeq 9.5$. In addition, we measure for the first time the redshift evolution of low-mass spin bias, up to $z=1.5$. Throughout this work, the ``low-mass range" is defined as the range of masses below the redshift-dependent {\it{crossover mass}} $\text{M}_{\text{c}}(z)$, i.e., the mass at which the spin bias signal inverts. Note that this mass is significantly smaller than the characteristic mass M$_*$.

In order to quantify the dependence of halo clustering on $\lambda$, we measure the relative bias, b$_{\lambda}$, of a $\lambda$-selected subpopulation following the standard procedure described in \cite{Sato-Polito2019}. We fix the primary bias property $\text{M}_{\text{vir}}$ and calculate, for a certain distance $r$, the ratio between the 2-point correlation functions of the $\lambda$-subpopulation and the entire population. The relative spin bias is therefore defined as:

\begin{equation}
   b^2_{\lambda}(r, \text{M}_{\text{vir}}) = \frac{\xi_\lambda(r, \text{M}_{\text{vir}})}{\xi(r, \text{M}_{\text{vir}})}.
   \label{eq:bias}
\end{equation}

Note that, in addition to the auto-correlation, we include the cross-correlation between different mass bins. The 2-point correlation function is measured using the \texttt{Corrfunc} code \citep{corrfunc2017} within the range of scales 5 $h^{-1}$Mpc $<r<$ 15 $h^{-1}$Mpc. Only distinct haloes (i.e., those whose centre does not lie within a larger halo) with more than 500 particle are included in the spin bias measurement and in all subsequent analysis of this paper. 

In order to estimate errors, the VSMDPL, SMDPL, and MDPL2 boxes are divided into equal-size subboxes of side lengths $L/2$, $L/3$ and $L/4$, respectively (where $L$ is the total size of each box). The resulting catalogues are further divided in log(M$_{\text{vir}}/h^{-1} \text{M}_\odot$) bins of width 0.15 dex. For each mass bin, the final value of the relative bias is the average over all subboxes and scales within the distance range considered. The errors correspond to the box-to-box standard deviation
(see \citealt{Sato-Polito2019} for more details).

Figure~\ref{spinbias} displays the redshift evolution of the relative bias for 25$\%$ $\lambda$-subsets (quartiles containing the higher- and lower-spin subpopulations). We have checked that the results are qualitatively the same if other percentiles are chosen. The evolution of spin bias is determined by the evolution of the crossover mass M$_{\text{c}}(z)$, which shifts towards lower masses as we move to higher redshifts: $\log(\text{M}_{\text{c}}/h^{-1} \text{M}_\odot )\simeq 11.5$ at $z=0$, and $\simeq 10$ at $z=0.8$. Above $z \sim 1$, the crossover is no longer evident,
so it is fair to say that the spin bias inversion is predominantly a late effect 
within the mass range considered. 
Figure~\ref{spinbias} also illustrates the advantage of including the VSMDPL box, which allows us to extend the mass range with respect to \cite{Sato-Polito2019} by more than 1 order of magnitude at the low-mass end.     

\begin{figure}
    \centering
    \includegraphics[scale=0.51]{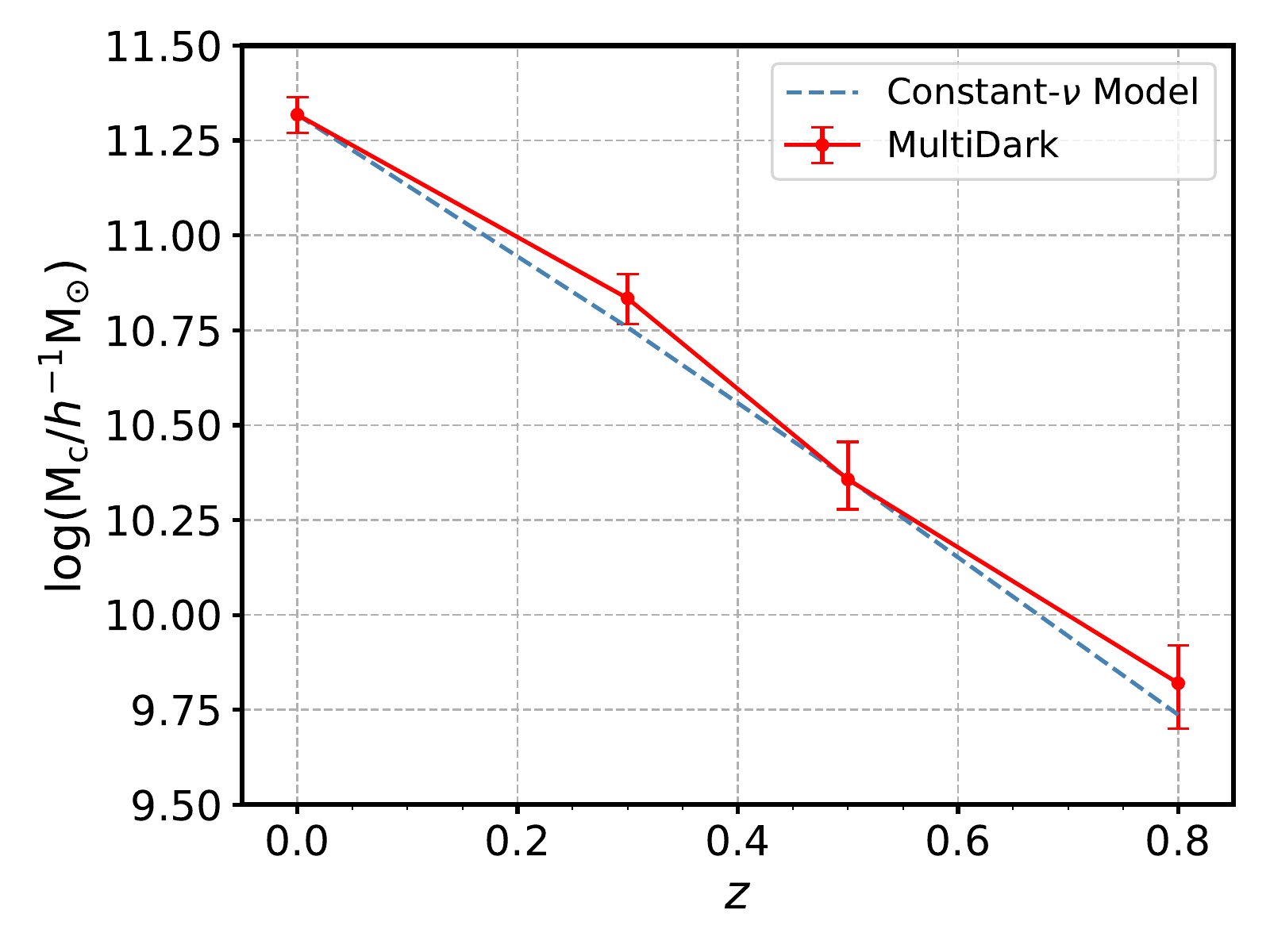}
    \caption{The redshift evolution of the spin-bias crossover mass, $\text{M}_{\text{c}}$, measured from MultiDark (red) and from a constant peak height model (blue) that assumes $\nu (\text{M}_\text{vir},z) = \delta_c/\sigma(\text{M}_\text{vir},z) = \nu (\text{M}_\text{c},0)$. The MultiDark data points and errors are computed from an MCMC procedure using data from Figure~\ref{spinbias}.}
    \label{cross}
\end{figure}

\begin{figure*}
    \centering
    \includegraphics[scale=0.58]{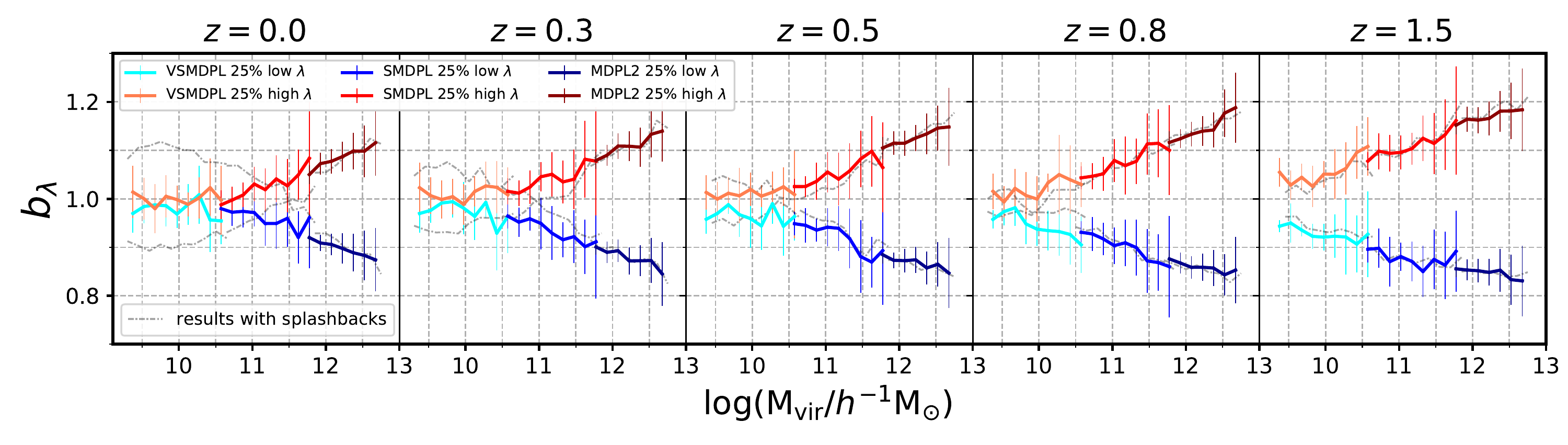}
    \caption{Same as Figure~\ref{spinbias} but removing splashback haloes, i.e., distinct haloes at the corresponding redshift which previously were subhaloes. Grey dash-dotted lines show the mean relative bias from Figure~\ref{spinbias} calculated with splashback haloes. As explained in the text, after their removal only the {\it{intrinsic}} mass dependence of spin bias remains at the low-mass end.} 
    \label{nosplashbacks}
\end{figure*}

A better way to visualise the evolution of the crossover mass is provided by Figure~\ref{cross}, where M$_{\text{c}}$ is plotted as a function of redshift. 
Here, M$_{\text{c}}(z)$ is computed from a simple Markov chain Monte Carlo (MCMC) procedure using the data shown in Figure~\ref{spinbias}. The evolution of the crossover mass is compared with a model that assumes constant peak height $\nu (\text{M}_\text{vir},z) = \delta_c/\sigma(\text{M}_\text{vir},z) = \nu (\text{M}_\text{c},0)$, which characterises the density fluctuations independently of redshift. 
The agreement between M$_{\text{c}}$ and the aforementioned model is excellent, implying that this mass scale
tracks the evolution of the large scale structure growth and, as a consequence, also the contribution from splashback haloes to the sample.

Finally, halo spin is known to be sensitive to particle resolution, which could potentially affect the lower-mass bins for each box (see \citealt{Benson2017}). For this reason, the 500-particle threshold is imposed. Note, however, that  VSMDPL haloes in the $\log(\text{M}_{\text{vir}}/h^{-1}\text{M}_\odot)= 10.5$ bin (where the inversion at $z=0$ is already observed) contain $\sim 5000$ particles, guaranteeing a precise measurement of halo spin and thus the robustness of our results. 
\section{The Effect of Splashback Haloes} 
\label{sec:splashback}

Splashback haloes are distinct haloes that were subhaloes at some previous time, i.e., passed through the virial radius of a larger halo.
Also referred to as ``ejected subhaloes", this population is known to be one of the main causes of assembly bias at the low-mass end (e.g., \citealt{dalal2008, Sunayama2016, Mansfield2020}). In this section, we show that splashback haloes also have a strong effect on spin bias. In order to identify this type of haloes in the MultiDark simulations at each redshift $z$, we select all distinct haloes (MultiDark \texttt{pid} $=-1$) that were previously subhaloes (MultiDark $z_{\text{firstacc}}>z$, where $z$ is the redshift under analysis and $z_{\text{firstacc}}$ is the first accretion redshift at which the main progenitor of the halo passed inside the virial radius of a larger one).

Most splashback haloes are low-mass haloes that are still near their previous host haloes, in the so-called ``splashback radius'', which covers a few times the virial radius of the host \citep{Ludlow2009, Wang2009, Adhikari2014, More2015}. Most of them are, therefore, still bounded to the potential wells of these massive haloes. Importantly, since splashbacks tend to live near their previous hosts, they trace these massive haloes on large scales and thus have a higher bias than other haloes of similar mass (e.g., \citealt{Wang2009}). As we discuss in Section~\ref{sec:tidal} below, their internal properties are also determined by tidal interactions with more massive haloes.

\begin{figure}
    \centering
    \includegraphics[scale=0.51]{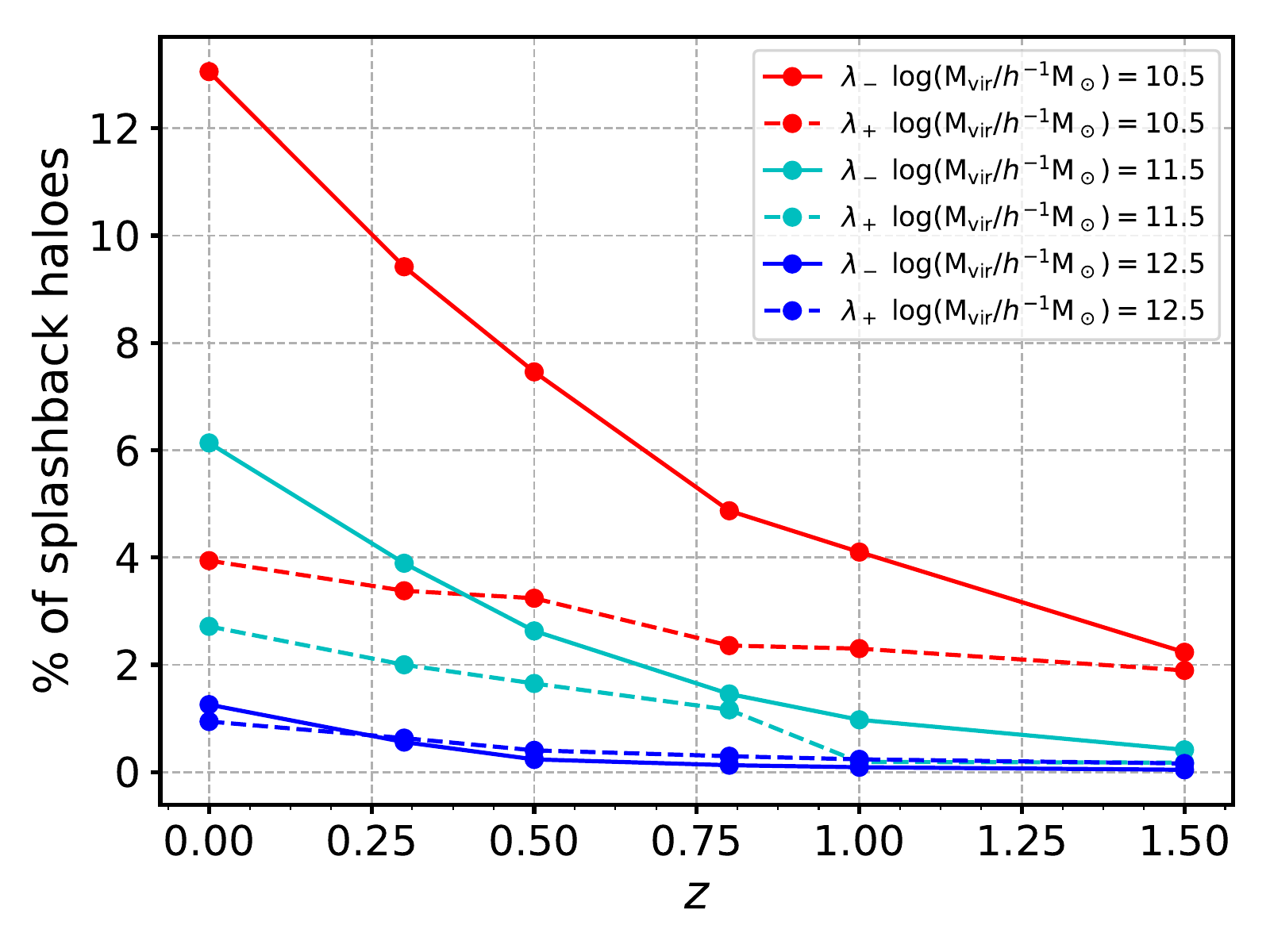}
    \caption{Percentage fraction of splashback haloes as a function of redshift for different mass bins of width 0.15 dex. Continuous and dashed lines show the 25\% lowest ($\lambda_-$) and  25\% highest ($\lambda_+$) spin subpopulations, respectively. Different colours represent different mass bins, each one calculated in the box where the relative spin bias is measured, i.e., $\log(\text{M}_{\text{vir}}/h^{-1}\text{M}_\odot)= 10.5$, 11.5 and 12.5 in VSMDPL, SMDPL and MDPL2, respectively.}
    \label{spredshift}
\end{figure}

\begin{figure*}
    \centering
    \includegraphics[scale=0.54]{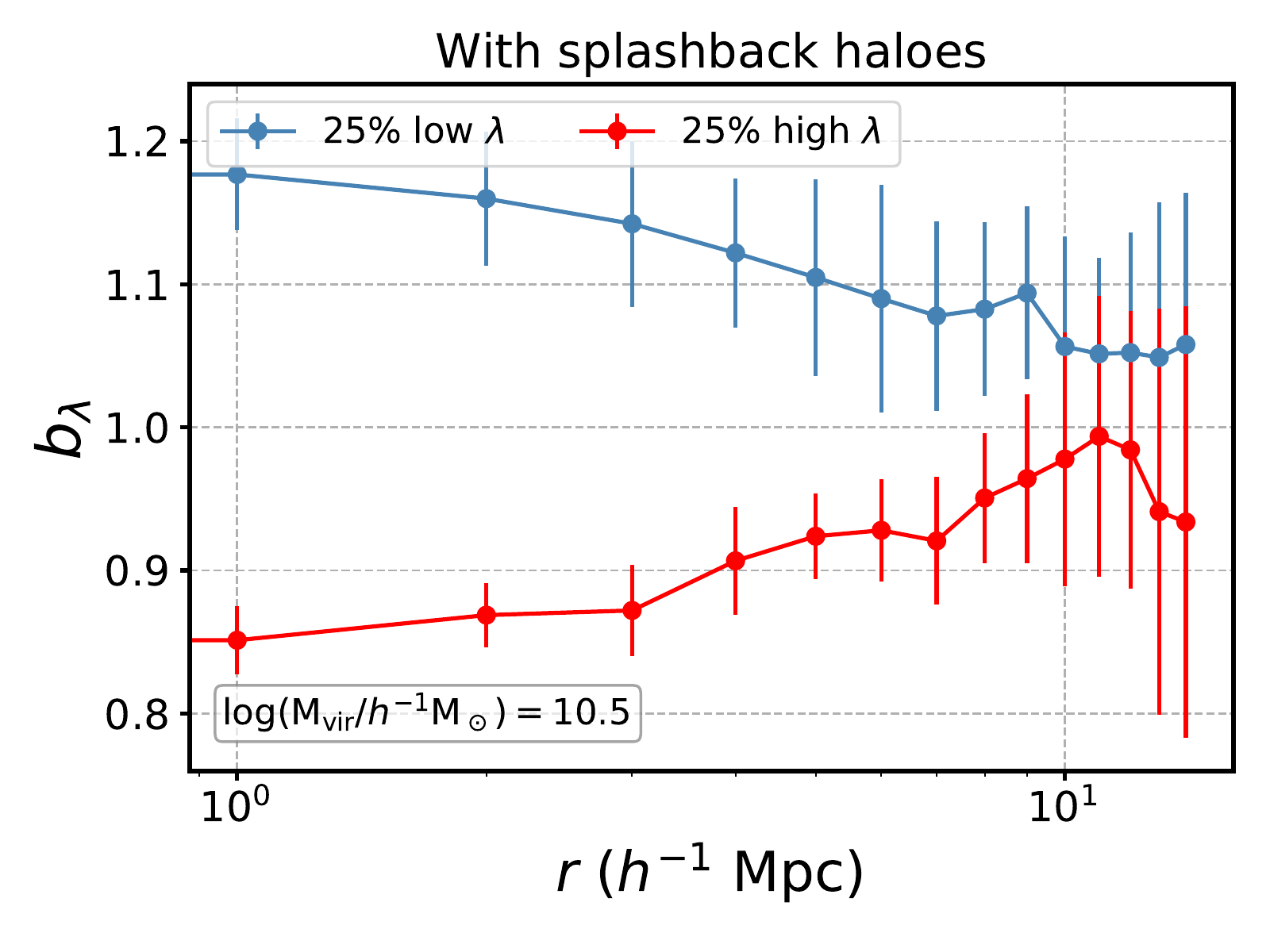}
    \includegraphics[scale=0.54]{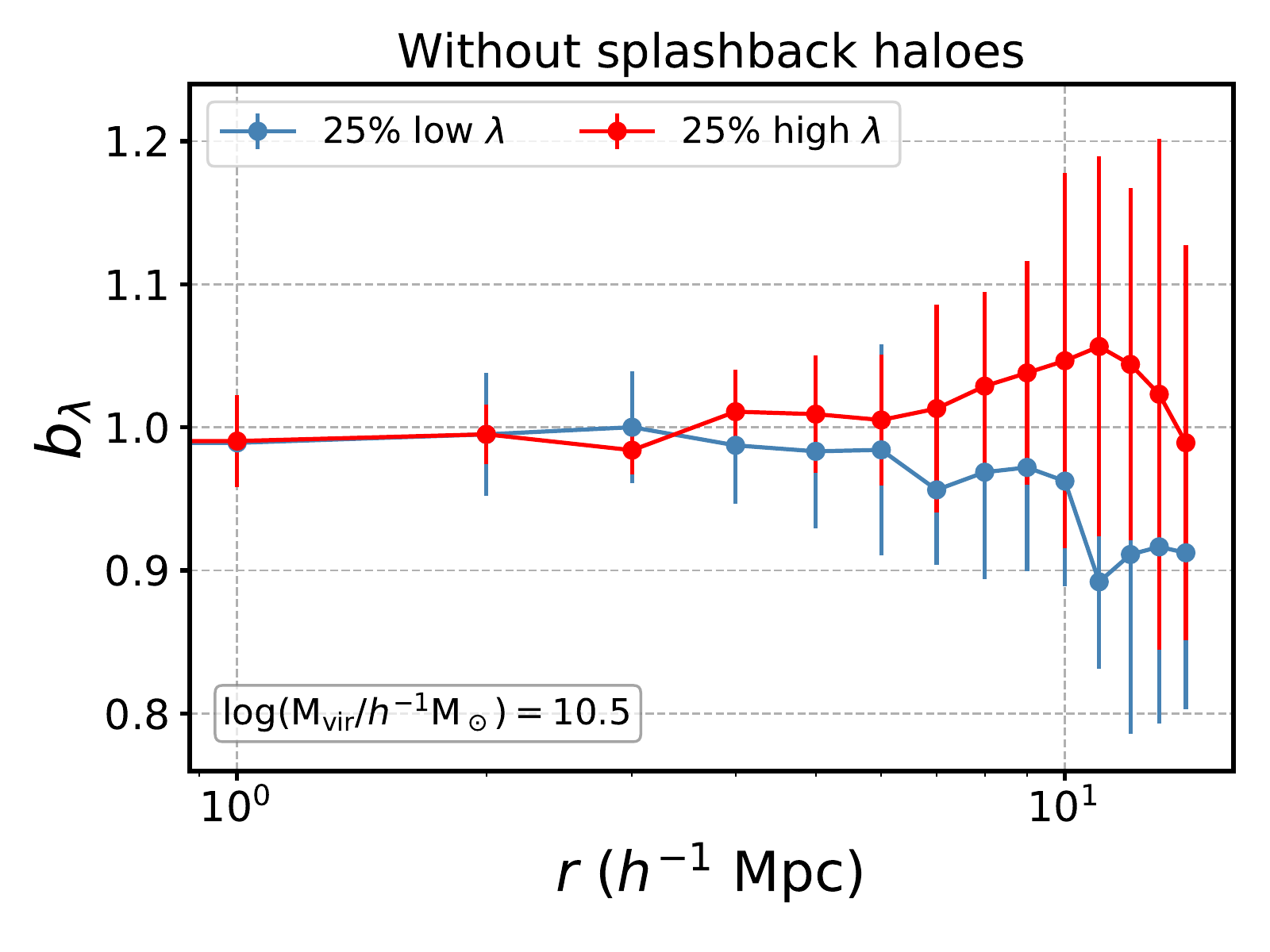}
    \caption{The scale dependence of low-mass spin bias. \textit{Left}: Relative spin bias as a function of scale for all haloes in a mass bin centred at $\log(\text{M}/h^{-1} \text{M}_\odot ) = 10.5$ (width 0.15 dex) in the VSMDPL box at $z=0$. \textit{Right}: Same but removing splashback haloes in order to show the cleaning of the signal towards small scales. Error bars show the box-to-box variation computed from a set of subboxes (see text).}
    \label{scale}
\end{figure*}

In Figure~\ref{nosplashbacks} we show, in the same format as in Figure~\ref{spinbias}, that removing the splashback population has a strong impact on spin bias at the low-mass end. Note that here the removal of splashbacks is performed prior to the $\lambda$-quartile definition, although an a posteriori removal of splashbacks does not change our results in any significant way. The most notable difference with respect to Figure~\ref{spinbias} is that the inversion of the signal for less massive haloes has completely disappeared. At $z=0$, the relative bias below $\log(\text{M}_{\text{vir}}/h^{-1} \text{M}_\odot )\simeq 10.5$ is consistent with zero, i.e., there is no difference between the clustering of low-spin and high-spin haloes when the population of splashback haloes is removed. As we move to higher redshifts, a residual amount of signal remains, but in a way that is consistent with the higher-mass trend. This clearly paints a picture in which the spin bias signal of Figure~\ref{spinbias} can be viewed as the result of two counteracting effects. First, an {\it{intrinsic}} effect that increases with halo mass and makes high-spin haloes be more tightly clustered than low-spin ones. Second, an opposite bias trend that is maximal for lower-mass haloes and is due to the presence of a very specific halo population. 

The magnitude of spin bias at the low-mass end is therefore determined by the amount of intrinsic signal present and the fraction of splashback haloes. Note that for the effect of splashbacks to be efficient, a mismatch in their abundances in the quartiles is expected. Figure~\ref{spredshift} displays the percentage of splashback haloes as a function of redshift in 3 narrow mass bins (of width 0.15 dex), in both the high- and low-$\lambda$ subsets. At $z=0$, the splashback fraction is very similar in the two quartiles for the halo mass bin centred at $\log(\text{M}_{\text{vir}}/h^{-1}\text{M}_\odot) = 12.5$. However, splashbacks become progressively more dominant in the low-$\lambda$ subset as we move to lower masses. Since splashback haloes ``carry" the high bias of their recent massive hosts, it is these lopsided fractions that causes the inversion of the signal seen in Figure~\ref{spinbias}.  Note that the inversion is still present if cross-correlations between different mass bins are not taken into account. However, the inclusion of cross-correlations does enhance the signal, since, unlike other haloes of similar mass, splashbacks are mostly found in high-density regions surrounded by massive haloes.

\begin{figure}
    \centering
    \includegraphics[scale=0.51]{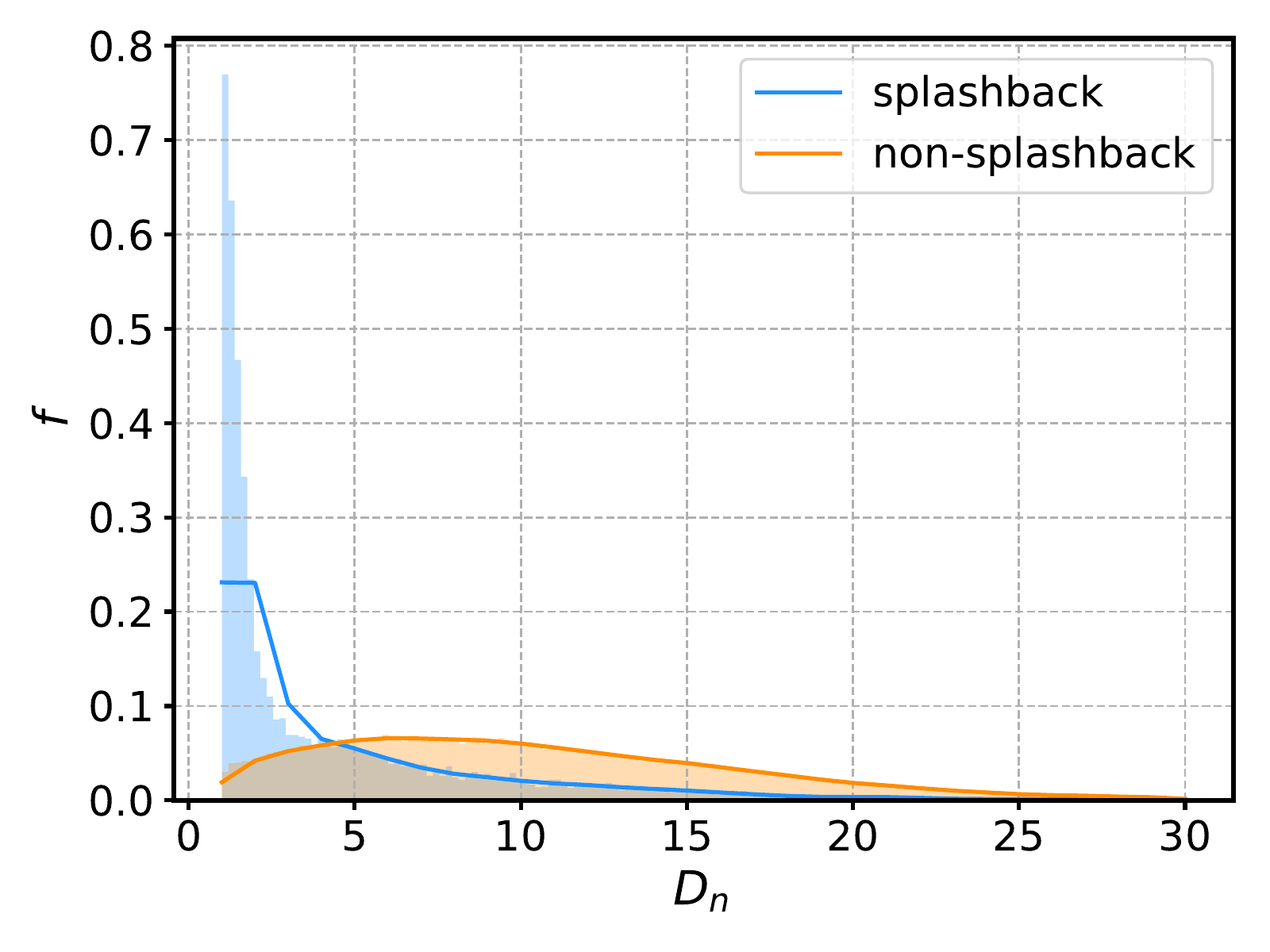}
    \caption{Normalised distributions and Gaussian kernel density estimators for the distance to a neighbour at least 300 times more massive ($D_n$ in units of the neighbour virial radius) for splashback and non-splashback haloes of $\log(\text{M}_\text{vir}/h^{-1} \text{M}_\odot ) = 10.5$  (mass width 0.15 dex) at $z=0$ in VSMDPL.}
    \label{Neighbour}
\end{figure}

\begin{figure*}
    \centering
    \includegraphics[scale=0.54]{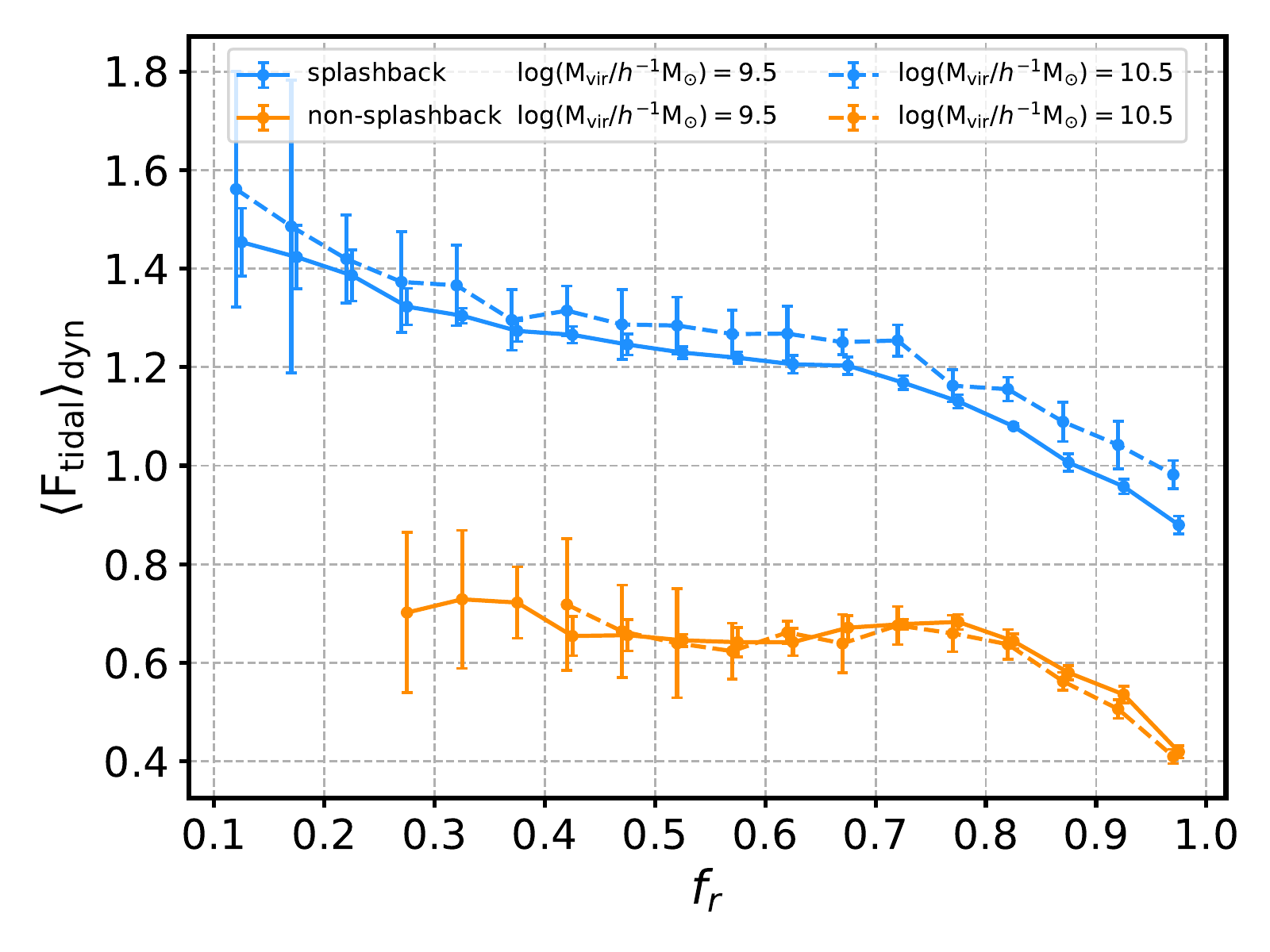}
    \includegraphics[scale=0.54]{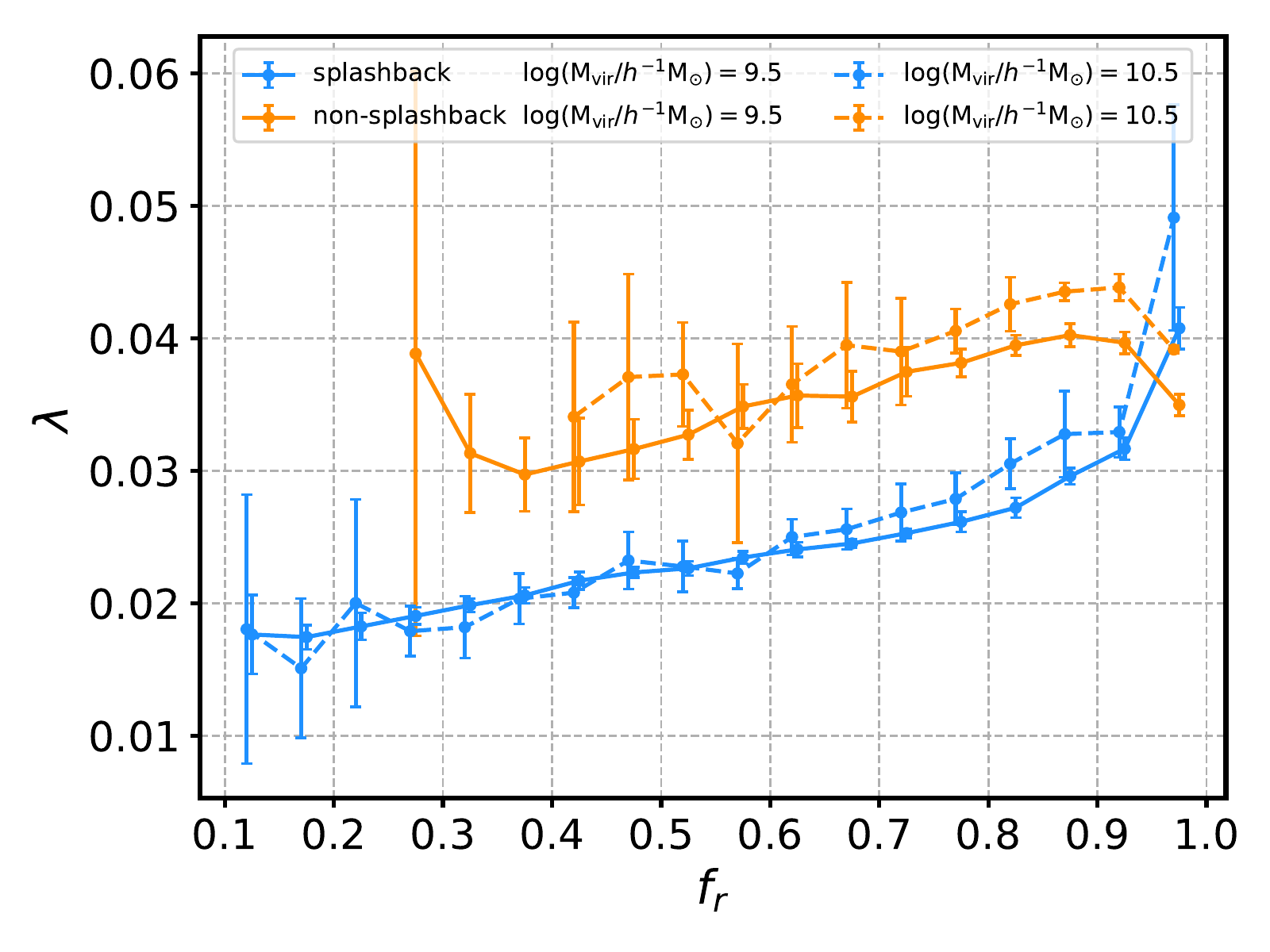}
    \caption{The effect of tidal interactions at $z=0$ in VSMDPL. \textit{Left}:  Mean dimensionless tidal force estimator, F$_{\rm tidal}$, averaged over past dynamical time, as a function of the retained mass fraction $f_r$ for splashback and non-splashback haloes in the mass bins centred at $\log(\text{M}_{\text{vir}}/h^{-1}\text{M}_\odot) = 9.5$ and $\log(\text{M}_{\text{vir}}/h^{-1}\text{M}_\odot) = 10.5$ (width 0.15 dex), respectively. The errors correspond to the standard deviation of mean time-averaged F$_{\rm tidal}$ evaluated in 8 subboxes. \textit{Right}:  Mean spin for the same subsets of haloes.}
    \label{Mpeak}
\end{figure*}

Figure~\ref{spredshift} also shows the redshift evolution of the splashback fractions. The evolution of the mass scale of crossover, M$_c$, is again determined by the difference in the abundance of splashbacks in the low-spin and high-spin quartiles for each mass bin. At $z=0$, this scale is $\log(\text{M}_{\text{c}}/h^{-1} \text{M}_\odot ) \simeq 11.5$, i.e., this is the mass at which the relative bias of each spin quartile is the same as that of the entire population. In Figure~\ref{spredshift} we can see that, at M$_c(z=0)$, the mismatch between splashback fractions is of a factor of $\sim$2. This is the amount of ``splashback signal" required to compensate for the intrinsic spin bias signal at that mass and redshift (note that once splashbacks are removed, the remaining relative spin bias ratio is $\sim$ 10$\%$ at M$_c(z=0)$, see Figure~\ref{nosplashbacks}). A similar behaviour is observed at $z=0.5$: a mismatch of a factor of $\sim$2 in the splashback fractions produces the cancellation of the spin bias signal at $\log(\text{M}_{\text{c}}(z=0.5)/h^{-1}\text{M}_\odot) \simeq 10.5$. Note that these abundance differences are important as long as a significant population of splashback haloes is present in the sample. At $z \gtrsim$ 1, this fraction is too small to produce any visible effect on the spin bias signal, as shows Figure~\ref{spinbias}.  

Another interesting aspect to investigate is the scale dependence of low-mass spin bias. In the left-hand panel of Figure~\ref{scale}, the relative spin bias is displayed as a function of scale for VSMDPL haloes in the mass bin centred at $\log(\text{M}/h^{-1}\text{M}_\odot) = 10.5$ (width 0.15 dex), at $z=0$. Note that here our measurements start from the scale $r=$ 1 $h^{-1}$Mpc instead of 5 $h^{-1}$Mpc. The presence of splashbacks not only produces the inversion of the spin bias signal but also enhances its amplitude on small scales. Removing this population (as shown in the right-hand panel) completely cleans off the signal on small scales, with only a very small intrinsic spin bias signal remaining on ``secondary bias scales" (i.e., $r\gtrsim$ 10 $h^{-1}$Mpc). Note that, in the context of assembly bias, the effect of splashback haloes is also more significant on small scales (e.g., \citealt{Sunayama2016}).

\section{Tidal Interactions} 
\label{sec:tidal}

In the previous Section, we show the impact that splashback haloes have on spin bias at the low-mass end, producing the inversion of the signal. This effect is a direct consequence of the low spin and high bias displayed by this population. In this section, we review one of the main physical mechanisms that gives rise to the splashback properties: the strong {\it{tidal interactions}} produced by large haloes (see, e.g., \citealt{Tormen1998,Kravtsov2004,Knebe2006,Wang2009,Hahn2009}).

Following \cite{salcedo2018}, in Figure~\ref{Neighbour} we show the distribution of the minimum distance to a massive neighbour, $D_n$, in VSMDPL at $z=0$, for splashback and non-splashback haloes separately, in units of the neighbour virial radii, R$_{\text{vir},n}$. A ``massive halo" in this context is defined as a distinct halo at least 300 times more massive than the corresponding halo under analysis. This computation is performed for haloes in a mass bin centred at $\log(\text{M}_{\text{vir}}/h^{-1}\text{M}_\odot)= 10.5$ (and width 0.15 dex). As expected, splashback haloes are, on average, significantly closer to a massive neighbour than other distinct haloes of the same mass, with $71\%$ of them lying within a distance of 5 R$_{\text{vir},n}$. In contrast, only $20\%$ of the non-splashback distinct population lie within the same distance to a massive halo. This, of course, explains the higher bias of these objects, since their spatial distribution correlates with that of massive haloes on relatively large scales.  

The proximity to a massive neighbour dictates that most of the splashback haloes are, to some extent, still bounded to the potential wells of their previous hosts. These haloes, therefore, are not only gravitationally influenced by their hosts during the subhalo phase, but also when they are outside its virial radius. Splashbacks are expected to be subject to extreme tidal forces, which tend to pull material away from their outer regions in a physical process called tidal stripping. For low-mass haloes, this mechanism not only results in mass loss, but also tends to reduce halo spin (see, e.g., \citealt{Lee2018}). Tidal stripping affects even more violently subhaloes (e.g., \citealt{Green2019}), which show lower spins than distinct haloes independently of the subhalo finder used \citep{Onions2013,Wang2015}.

In order to investigate whether the lower spin of splashback haloes is caused by tidal stripping, we
follow a similar procedure as \cite{Lee2018} and evaluate its effect on the spin parameter in VSMDPL at $z=0$. For this, we define the fraction of mass {\it{retained}} by a halo after it reaches its peak mass, M$_{\text{peak}}$, as $f_{\rm r}=$ M$_{\text{vir}}$/M$_{\text{peak}}$ (where M$_{\text{peak}}$ is the maximum mass that a halo reaches throughout its entire accretion history). We complement this analysis using the tidal force parameter, F$_{\rm tidal}$, provided by ROCKSTAR. F$_{\rm tidal}$ is simply defined as the ratio  R$_{\text{vir}}$/R$_{\text{Hill}}$, where R$_{\text{Hill}}$ is the ``Hill radius", which can be expressed as:

\begin{equation}
    \text{R}_{\text{Hill}}\simeq d \, \left(\frac{m}{3M} \right)^{1/3}
\end{equation}

\noindent and represents the {\it{sphere of influence}} of a halo. In essence, in the restricted three-body problem, consisting of a body of mass $M$ at a distance $d$ from a smaller body of mass $m \ll M$, a third body of negligible mass can have stable circular orbits around the smaller mass $m$ only within the Hill radius (see \citealt{MurrayDermott1999} for more details). It is noteworthy that haloes in general do not follow circular orbits around each other and that F$_{\rm tidal}$ only considers the strongest tidal force from any nearby halo. This is sufficient, however, for the purposes of this test, since for splashbacks and subhaloes their (previous) host halo is in general the most tidally influential nearby halo. A halo is prone to be subject to significant tidal stripping when F$_{\rm tidal}$ is typically greater than 1 (e.g., \citealt{Hahn2009, Lee2018}).

The effect of tidal interactions becomes clear in Figure~\ref{Mpeak}, where we plot the mean F$_{\rm tidal}$ averaged over past dynamical time and the mean $\lambda$ as a function of $f_{\rm r}$, for splashback and non-splashback haloes separately. The left-hand panel of Figure~\ref{Mpeak} displays a clear anti-correlation between the mass retained by a halo and the tidal force that it suffers. It also shows the significantly higher tidal forces experienced by splashbacks as compared to other distinct haloes. This is a clear sign that the mass loss of splashbacks is mostly due to tidal stripping (note that the values of the mean F$_{\rm tidal}$ are almost always above 1 for splashback haloes). The right-hand panel of Figure~\ref{Mpeak}, on the other hand, illustrates the effect on spin: 
the more severe the mass loss, the lower the spin of the haloes. In both panels, non-splashback haloes never reach extremely low $f_{\rm r}$, which shows that the mass loss of splashback haloes is more significant. These results, which are in agreement with previous findings (see, e.g., \citealt{Lee2018}), highlight the connection between spin and tidal stripping for splashback haloes. 

Finally, some theories relate lower spin parameters with halo histories characterised by the number of major merger events \citep{Vitvitska2002}. Lower spin values for splashbacks could also potentially be caused by a lack of major merger events during their lifetimes. However, we have checked that the fraction of recent major mergers ($z<3$) is approximately the same for splashback and non-splashback distinct haloes in VSMDPL, favouring the tidal stripping mechanism. These mechanisms need, however, further investigation that we will address in follow-up work. 
\section{Discussion \& Conclusions} 
\label{sec:conc}

In this paper, we measure the redshift evolution and scale dependence of halo spin bias for low-mass haloes, in order to investigate the physical origins of the effect. Our main results can be summarised as follows:

\begin{itemize} 
    \item We confirm the existence of the low-mass spin bias inversion, which makes low-spin haloes cluster more strongly than high-spin haloes below the crossover mass, $\text{M}_{\text{c}}$. This result is in agreement with previous findings by \cite{Sato-Polito2019} and \cite{Johnson2019}. The VSMDPL box allows us to reach a range of small masses that has never been probed before, i.e., $\log(\text{M}_{\text{vir}}/\text{M}_\odot h^{-1})\lesssim9.5$. 
    \item The redshift evolution of spin bias is determined by the location of $\text{M}_{\text{c}}$, which shifts towards lower masses as we move to higher redshift. This signal crossover falls outside our mass range at $z\sim1$  and its evolution follows that of a constant peak-height model.  
    \item We show that excluding splashback haloes has a major impact on spin bias at the low-mass end, since it completely removes the inversion of the signal at all redshifts (this is particularly evident on small scales). The small signal that remains follows perfectly the high-mass trend, by which higher-spin haloes are progressively more tightly clustered than lower-spin haloes.   
    \item The crossover arises in the mass scale where the fraction of splashback haloes is sufficiently larger in the low-spin quartile than in the high-spin subset. The unbalance between spin quartiles makes the inversion show up only at the low-mass end and at recent times, since the fraction of splashbacks is only significant at low-masses and low redshifts.
    \item We confirm using the VSMDPL box that splashback haloes of masses below the crossover are most likely to be found near massive neighbours (of at least 300 times their own mass). This explains their high bias compared to other distinct haloes of the same mass, as previously reported in the context of assembly bias studies (see, e.g., \citealt{dalal2008, salcedo2018, Mansfield2020}).
    \item We further show that, in the VSMDPL box, splashbacks have experienced significantly stronger tidal forces than other distinct haloes of similar mass, which points towards tidal stripping from massive neighbours as the main mechanism behind their typically low spin.
\end{itemize}

In the light of our findings, spin bias can be viewed as the result of two separate effects. On the one hand, the {\it{intrinsic spin bias}}, which dominates the high-mass end and dictates that higher-spin haloes of a given mass have higher bias. On the other hand, the counter-acting {\it{splashback effect}}, which imposes an opposite bias trend and is only relevant for low-mass haloes. The removal of splashback haloes can thus be seen as a sort of  ``cleaning" of the signal, since the inversion of the trend, which is more pronounced at the low-mass end and at small scales, completely disappears. 

Our results align well with previous works that have shown certain degree of connection between different secondary bias dependencies; in particular, spin and assembly bias \citep{Sato-Polito2019,Johnson2019}. By revealing the effect that splashback haloes have on the former, we found a common mechanism that links both effects. In this context, we have confirmed that removing splashbacks also reduces the amplitude of low-mass assembly bias, as previously reported by, e.g., \cite{dalal2008,Wang2009,Sunayama2016,Mansfield2020}. 
In the range of small masses covered by VSMDPL, however, this only accounts for $\sim$1/3 of the total signal (see Appendix~\ref{sec:app}). In comparison, this result emphasises the importance of splashbacks for spin bias, where their impact is clearly more significant.

As part of our follow-up work, we will explore the the role of tidal anisotropy as a mediator in the connection between internal halo properties and the large-scale halo bias (see, e.g., \citealt{Ramakrishnan2019, Paranjape2020}). We are interested in linking these ideas to models in which the halo angular momentum is induced by the surrounding tidal field, with the aim of shedding light onto the physical origins of the aforementioned ``intrinsic'' spin bias dependence. Note that many works have focused on the mechanisms that make haloes acquire spin. Some potential explanations are based on the tidal torque theory (TTT, e.g., \citealt{Peebles1969, Doroshkevich1970, White1984, BarnesEfstathiou1987}), in which the angular momentum is induced by the large-scale tidal field. There is, however, ample evidence for deviations from TTT, especially in recent times (see, e.g., \citealt{Porciani2002, Lopez2019}). Another possibility is that this acquisition of angular momentum takes place through a random-walk process as haloes accrete satellites \citep{Vitvitska2002}. 

An interesting phenomenon that we also plan to investigate is the low spin of splashback haloes. As discussed before, this effect is likely to be caused by intense tidal stripping, a mechanism that typically produces both mass and spin loss. One possible reason for the spin loss is the notion that tidal stripping typically removes the outer layers of haloes, which typically contain higher angular momentum particles \citep{Lee2018}. Other possible explanations are related to ``dynamical self-friction'', where tidally stripped particles torque the remaining material and make it lose specific orbital energy \citep{Miller2020}.

Finally, as showed in \cite{Montero-Dorta2020b}, spin bias provides an alternative route towards an observational detection of secondary bias (see previous attempts in the context of galaxy assembly bias in, e.g., \citealt{miyatake2016,Lin2016,MonteroDorta2017, niemiec2018}). The intrinsic halo spin bias signal, unlike assembly bias, is expected to be maximal for the most massive (and thus easier to detect) clusters. Although measuring the spin of haloes is still challenging with current instrumentation, upcoming surveys such as the Square Kilometer Array (SKA\footnote{https://www.skatelescope.org}) and observational techniques such as the {\it{kinetic Sunyaev Zel'dovich}} effect  \citep{SZ1970,SZ1980A,SZ1980B} will provide exciting opportunities in the future  (see discussion in, e.g., \citealt{Mroczkowski2019}). 

\section*{Acknowledgments}

BT and ADMD thank FAPESP for financial support. LRA thanks both FAPESP and CNPq for financial support. MCA acknowledges financial support from the Austrian National Science Foundation through FWF stand-alone grant P31154-N27.

We are very thankful for the constructive comments provided by the referee Andres Salcedo. BT thanks Marcos Lima for valuable suggestions during first stages of this work. We also thank Nelson Padilla, Facundo Rodriguez, and Raul Angulo for fruitful discussions during the I Mini Workshop on Assembly Bias held in Santiago, Chile. 

The CosmoSim database used in this paper is a service by the Leibniz-Institute for Astrophysics Potsdam (AIP).
The MultiDark database was developed in cooperation with the Spanish MultiDark Consolider Project CSD2009-00064.
The authors gratefully acknowledge the Gauss Centre for Supercomputing e.V. (www.gauss-centre.eu) and the Partnership for Advanced Supercomputing in Europe (PRACE, www.prace-ri.eu) for funding the MultiDark simulation project by providing computing time on the GCS Supercomputer SuperMUC at Leibniz Supercomputing Centre (LRZ, www.lrz.de).
The Bolshoi simulations have been performed within the Bolshoi project of the University of California High-Performance AstroComputing Center (UC-HiPACC) and were run at the NASA Ames Research Center.

\section*{Data Availability}

The simulation data underlying this article are publicly available at the CosmoSim website (see Sections \ref{sec:intro} and \ref{sec:sims}). No new data were generated in support of this research.

\bibliography{references}

\appendix

\section{Spin and Assembly Bias}
\label{sec:app}

In this appendix, we compare the impact of splashback haloes on spin bias and assembly bias at the low-mass end. To this purpose, we calculate the relative bias with the same procedure described in Sections \ref{sec:inversion} and \ref{sec:splashback} for halo concentration and age, both also computed from MultiDark. We define concentration as:
\begin{equation}
    c_{200} = \frac{\text{R}_{200}}{\text{R}_s},
\end{equation}
where $\text{R}_s$ is the Klypin scale radius \citep{klypin2011} and $\text{R}_{200}$ is given by:
\begin{equation}
    \text{M}_{200} = 200\,\rho_{\text{cr}}\,\frac{4 \pi}{3}\,\text{R}^3_{200}.
\end{equation}
Halo age, denoted as $a_{1/2}$, is the scale factor at which half of the peak mass of the halo was accreted.

\begin{figure}
    \centering
    \includegraphics[scale=0.95]{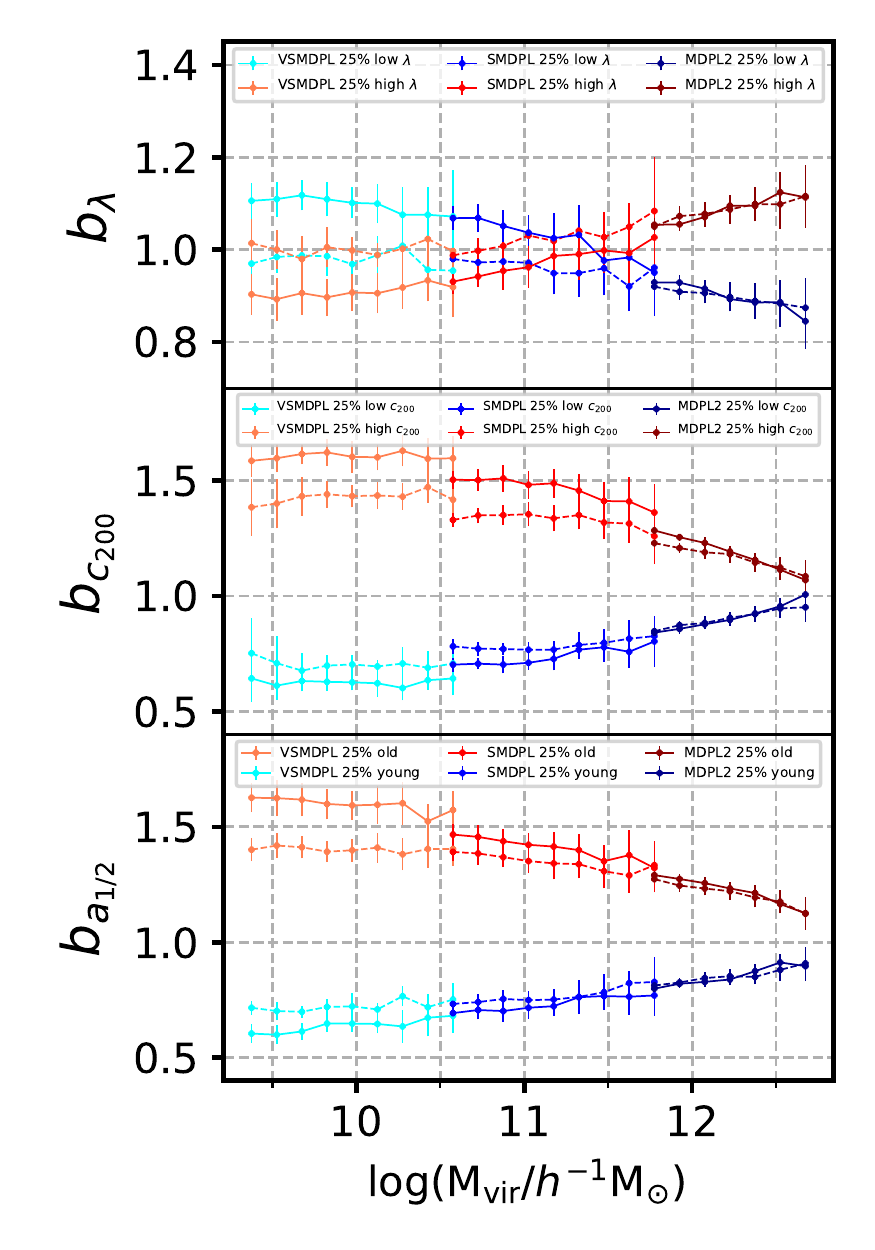}
    \caption{The effect of splashback haloes on spin bias and assembly bias (i.e., the secondary bias for halo concentration and age) at $z=0$ in VSMDPL, SMDPL, and MDPL2. Solid (dashed) lines display the relative bias measured by including (removing) splashback haloes.}
    \label{speff}
\end{figure}

Figure~\ref{speff} displays the relative bias measured at $z=0$ with and without splashback haloes (in solid and dashed lines, respectively) for spin, concentration, and age. While in VSMDPL, removing splashback haloes only reduce the assembly bias signal by $\sim30\%$, their impact on spin bias is far more severe, reaching 100\% and thus eliminating the effect completely (as discussed in Section~\ref{sec:splashback}). At high masses, where the percentage of splashback haloes becomes progressively negligible, their impact vanishes for all secondary bias trends.

\bsp
\label{lastpage}

\end{document}